\documentclass[aps,pra,twocolumn,showpacs,amsmath,amssymb]{revtex4-1}

\usepackage{amsmath}
\usepackage{amssymb}
\usepackage{graphicx}
\usepackage{color}
\usepackage{amsmath}
\usepackage{amssymb}

\begin{document}

\title{Unconventional quantum Hall effect in Floquet topological insulators}
\author{M. Tahir$^{1,\ddagger}$ P. Vasilopoulos$^{1,\dag}$, and U. Schwingenschl\"{o}gl$^{2,}$}
\email{udo.schwingenschlogl@kaust.edu.sa}
\affiliation{$^{1}$Department of Physics, Concordia University, Montreal, Quebec, Canada H3G 1M8}
\affiliation{$^{2}$King Abdullah University of Science and Technology (KAUST),
Physical Science and Engineering Division (PSE), Thuwal 23955-6900, Saudi Arabia}

\begin{abstract}
We study an unconventional quantum Hall effect for the surface states 
of ultrathin Floquet topological insulators in a 
perpendicular magnetic field. The resulting band structure is modified by photon dressing 
and the topological property is governed by the low-energy dynamics of a single surface.
An exchange of  symmetric and antisymmetric surface states occurs
by reversing the light's polarization. We find a novel quantum Hall state 
in which the zeroth Landau level undergoes a phase transition from 
a trivial insulator state, with Hall conductivity $\sigma_{yx}=0$ at zero Fermi energy, to a Hall insulator
state with $\sigma_{yx}=e^2/2h$. These findings open new possibilities 
for experimentally realizing nontrivial quantum states and unusual 
quantum Hall plateaux at $(\pm1/2,\pm3/2,\pm5/2,...)e^2/h$.
\end{abstract}

\maketitle


In the regime of the integral quantum Hall effect (IQHE) for conventional, 
two-dimensional (2D) systems, e.g., in a GaAs/AlGaAs heterostructure, 
the Hall conductivity takes the values 
$2(n+1) e^{2}/h=( 2, 4, 6,... )\ e^{2}/h$, where $h$ is the Planck constant, $e$ the electron 
charge, and $n$ an integer. In graphene though the IQHE plateaux appear at 
$4(n+1/2) e^{2}/h=(\pm 2, \pm 6, \pm 10,... )\ e^{2}/h$ \cite{KAS}, and 
the "half integer" aspect is hidden under the 4-fold degeneracy associated with the spin and valley 
degrees of freedom \cite{VPS}. More recently, the IQHE has been assessed for 
silicene \cite{MTU} and MoS$_2$ \cite{XCG} in which the spin-orbit interaction 
rearranges, as in 2D systems \cite{XF}, the Landau levels (LLs) 
in two groups and the plateaux appear at integer values of $ e^{2}/h$ due to a double degeneracy ($ \pm 0, \pm 1, \pm 2, \pm 4, \pm 6,...$). 
In topological insulators (TIs) electrons on both top and bottom surfaces
contribute to the Hall conductivity and its plateaux, due to the surface degeneracy, have heights $2(n+1/2) \ e^{2}/h =( \pm 0, \pm 1, \pm 3, \pm 5,... )e^{2}/h$ \cite{DY,HL,CC}. Though the quest for a genuine "half-integer" QHE is long, a QHE like $(n+1/2) \ e^{2}/h =(\pm 0, \pm 1/2, \pm 3/2, \pm 5/2,...)\ e^{2}/h$ {\it without any degeneracy prefactor},  has not 
been observed. Then one wonders whether a "half-integer" 
QHE is possible in TIs by breaking their surface degeneracy. 

TIs, well established theoretically \cite{MC} and experimentally \cite{MS}, are a state of matter,
that cannot appear in normal 2D systems with time-reversal symmetry \cite{BT}. 
They exhibit exotic properties such as disorder-protected conducting surface states, a single
Dirac cone, quantum phase transitions \cite{HYY,JGA}, etc. These findings generated a strong interest in TIs 
that was further intensified by their potential applications in quantum computing \cite{AJ}, optical 
devices \cite{OE}, terahertz detectors \cite{XJ}, etc.

More recently, the surface states of TIs driven by circularly polarized \textit{off-resonant} light 
have become a subject of strong interest \cite{NG,TT,ANA,HDT}. 
TIs driven by external time-periodic perturbations are known as Floquet TIs (FTIs). For such systems it is
convenient to use the Floquet theory \cite{TT}. 
In the appropriate frequency regime the {\it off-resonant} light cannot generate real photon absorption or emission due to energy conservation. Accordingly, it does not directly excite electrons but instead modifies the electron band structure through second-order virtual-photon absorption processes.
Averaged over time these processes result in an effective static alteration of the band structure.
Illuminating, e.g., graphene or silicene with \textit{off-resonant} light
generates a Haldane-type gap \cite{ZMA}.

Floquet bands were first realized in photonic crystals \cite{MJ} and have
been verified by recent experiments on the surface states of 
FTIs \cite{YH,HJJ}. These first studies of \textit{off-resonant} light were limited to the 
band structure of FTIs and differ from many optical effects in TIs
\cite{OE}. Also, no magnetic field was involved in these experiments. 

In this work we identify a novel quantum Hall state  of ultrathin FTIs in a  
magnetic field when their surface degeneracy is broken due to an \textit{off-resonant} light. We 
evaluate their band structure and the longitudinal and Hall conductivities
using linear response theory \cite{MK,PV}. 

\textit{Model formulation}. We consider surface states of ultrathin TIs in the ($x,y$) plane in the 
presence of circularly polarized \textit{off-resonant} light \cite{TT} and 
hybridization \cite{HZ} between the top and bottom surface states. Extending 
the 2D Dirac-like Hamiltonian \cite{TT,MTVP} by including an external 
perpendicular magnetic field $B$ gives
\begin{equation}
H_{s}^{l}=v_{F}(\sigma _{x}\mathbf{\Pi }_{y}-\sigma _{y}\mathbf{\Pi }
_{x})+s\Delta _{h}\sigma _{z}+l\Delta _{\Omega }\sigma _{z}, \label{1}
\end{equation}
where $s = +/-$ is for symmetric/antisymmetric surface states, $l=+/- $ 
for right-/left-handed circularly polarized \textit{off-resonant} light, 
($\sigma _{x}$, $\sigma _{y}$, $\sigma _{z}$) the Pauli
matrices and $v_{F}$ the Fermi velocity.
$\Delta _{h}$ is the hybridization energy between the top and bottom surface
states that, depending on the thickness, varies from 20 meV to 120 meV
\cite{YK}. 
$\Delta _{\Omega }=e^{2}v_{F}^{2}\hslash ^{2}A_{0}^{2}/\hslash
^{3}\Omega $ is the mass term induced by the \textit{off-resonant} light
with amplitude $E_{0}$, $\Omega $ the light's frequency, and $A_0=E_{0}/\Omega $. 
It breaks the time-reversal symmetry and its value is about
50 meV \cite{YH,HJJ}. $\mathbf{\Pi =p}+e\mathbf{A}$ is the 2D canonical
momentum with vector potential $\mathbf{A}$.
In the Landau gauge $\mathbf{A}=(0, Bx, 0)$, diagonalizing the Hamiltonian (1) gives
the eigenvalues
\begin{equation}
E_{n,s}^{\lambda ,l}=\lambda [\hslash ^{2}\omega _{c}^{2}n+\Delta
_{s,l}^{2}]^{1/2},\quad E_{0,s}^{0,l}=-\Delta _{s,l}, \label{2}
\end{equation}
where $\lambda =\pm 1$ represents the electron/hole states, $\omega _{c}=v_{F}
\sqrt{2eB/\hslash }$, and $\Delta _{s,l}=l\Delta _{\Omega }+s\Delta _{h}$. 
The corresponding normalized eigenfunctions are 
\begin{equation}
\hspace*{-0.35cm}\Psi _{n,s}^{\lambda ,l}=\frac{e^{ik_{y}y}}{\sqrt{L_{y}}}\Big( 
\begin{array}{c}
C_{n,s}^{\lambda ,l}\phi _{n-1} \\ 
D_{n,s}^{\lambda ,l}\phi _{n}
\end{array}
\Big),\,\,\,\Psi _{0,s}^{0,l}=\frac{e^{ik_{y}y}}{\sqrt{L_{y}}}\Big( 
\begin{array}{c}
0 \\ 
\phi _{0}
\end{array}
\Big), \label{3}
\end{equation}
where $C_{n,s}^{\lambda ,l}=[(E_{n,s}^{\lambda ,l}+\Delta
_{s,l})/2E_{n,s}^{\lambda ,l}]^{1/2}$, $D_{n,s}^{\lambda ,l}=
[(E_{n,s}^{\lambda ,l}-\Delta _{s,l})/2E_{n,s}^{\lambda ,l}]^{1/2}$; 
$\phi_{n}$ are the harmonic oscillator functions. 
Notice that Eq.\ (1) does not contain the Zeeman term $gs_zB$. We neglect
it, because we consider only weak $B$ fields $\leq 1$ T, cf.\ Fig.\ 1. For a $g$ factor
as large as 20 the Zeeman energy at $B=1$ T is $0.58$ meV and much smaller than all
other energies.

A close inspection of
Eq.\ (2) shows that we have a gapped Dirac spectrum, with gap $\Delta _{h}$, and electron-hole
symmetry for zero \textit{off-resonant} light, $\Delta _{\Omega }=0.$ For 
$\Delta _{\Omega }>\Delta _{h}>0$,
the eigenvalues of Eq.\ (2) show a $\sqrt{B}$ dependence and
the LLs split, see Fig. 1. We find an exchange of the
symmetric (solid curves) and antisymmetric (dotted curves) surface states
by changing the light's polarization from right (red curves) to left (black curves). 
The parameters used are $v_{F}=0.5\times 10^{6}$ m/s, $\Delta _{h}=20$ meV, and
$\Delta _{\Omega }=30$ meV ($ev_{F}A_{0}=0.48$ eV, $\hslash \Omega =7.5$ eV) \cite{TT}. 
The $n=0$ LL appears in the hole band for right-handed light 
(red curves, $l=1$) and in the electron band for left-handed 
light  (black curves, $l=-1$), see Fig.\ 1. The exchange of surface states
induced by the field $B$ and the \textit{off-resonant} light in 
such FTIs is an entirely new phenomenon. The
energies of the two surfaces are different for $B\rightarrow 0$, since the
gap $l\Delta _{\Omega } \pm \Delta _{h}$ increases for one surface and decreases for the other. 

We emphasize that the band gaps $l \Delta _{\Omega }+s \Delta _{h}$ at the two surfaces
of ultrathin FTIs can be made different by, e.g., varying the light's 
frequency or amplitude. This can create, e.g., for $l=1$, one surface
with a small gap $ \Delta _{\Omega }- \Delta _{h}$ and the other one with a large gap $ \Delta _{\Omega }+ \Delta _{h}$; for $l=-1$ these gaps could be exchanged. Accordingly, only the antisymmetric or 
symmetric surface contributes to the transport properties 
depending on the light polarization ($ l=\pm 1$). Such a situation could be realized 
in experiments on FTIs, similar to those of Refs. \cite{YH,HJJ}, by varying the sample thickness \cite{YK} down to 
the limit of 6 nm below which different
energies $\Delta_h$ have been reported \cite{YK}.
To our knowledge this is a novel state of matter
in FTIs like Bi$_{2}$Se$_{3}$, Bi$_{2}$Te$_{3}$, HgTe, and related materials.

\begin{figure}[t]
\includegraphics[width=0.8\columnwidth,height=0.45\columnwidth]{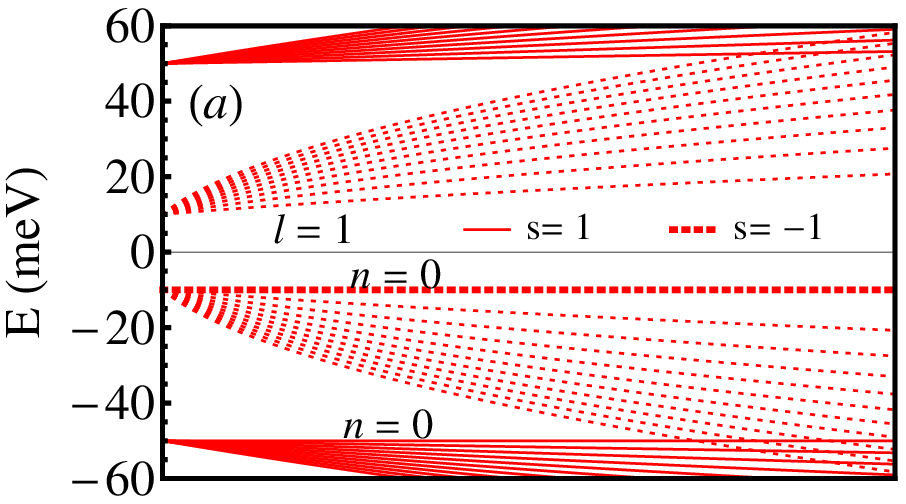}\vspace{-0.4 cm}
\hspace*{0.3 cm}\includegraphics[width=0.84\columnwidth,height=0.5\columnwidth]{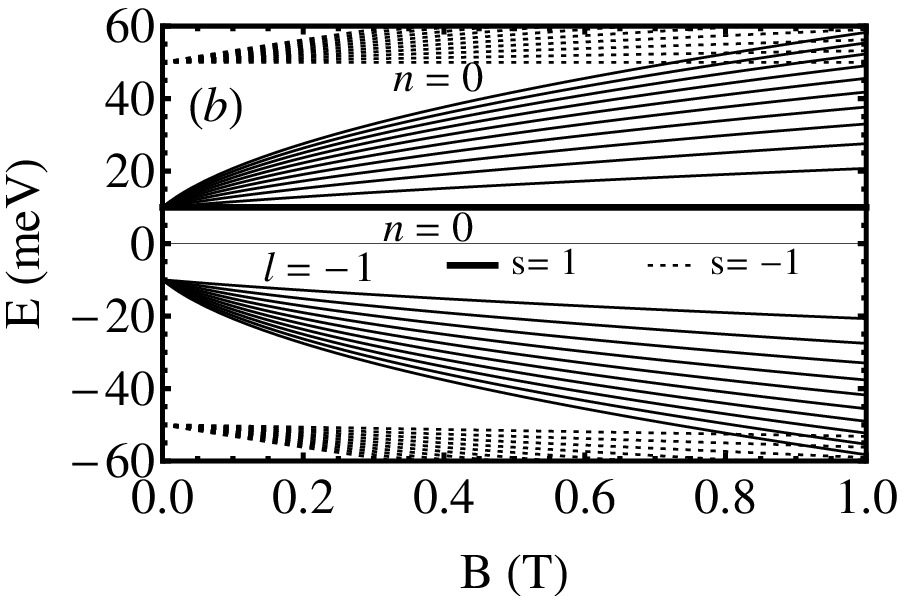}
\vspace*{-.25cm}
\caption{Energy spectrum versus magnetic field $B$. The symmetric (solid curves) and antisymmetric 
(dashed curves) surface states are shown for right-handed 
 (a) and (b) left-handed  light with $\Delta _{h}=20$ meV and
$\Delta _{\Omega }=30$ meV. Note that the first two $n=0$, $B$-independent LLs (thick lines) are
 in the valence (conduction) band for right-handed (left-handed) light.} 
\end{figure}

\textit{Longitudinal conductivity}. 
For weak scattering potentials  the current is due to hopping between orbit centres
as a result of carrier collisions with, e.g., charged 
impurities \cite{MK}.  In a normal magnetic field the diffusive contribution 
$\sigma_{xx}^{dif}$ to $\sigma_{xx}$ vanishes and only the collisional
contribution $\sigma_{xx}^{col}\equiv\sigma_{xx}$ is important; it is given by \cite{MK,PV}
\begin{equation}
\sigma _{xx}=\frac{e^{2}\beta }{2S_{0}}\sum_{\zeta \neq \zeta ^{\prime
}}f(E_{\zeta })[1-f(E_{\zeta ^{\prime }})]W_{\zeta \zeta ^{\prime
}}(X_{\zeta }-X_{\zeta ^{\prime }})^{2}, \label{4}
\end{equation}
where $f(E_{\zeta })=(\exp [\beta (E_{\zeta }-E_{F})]+1)^{-1}$ is the Fermi
Dirac distribution function, $\beta=k_{B}T$, $T$ the temperature, $k_{B}$ the Boltzmann
constant, and $E_{F}$ the Fermi energy.
$W_{\zeta \zeta ^{\prime }}$ is the transition rate between the
one-electron states $\left\vert \zeta \right\rangle $ and $\left\vert \zeta
^{\prime }\right\rangle $, and $e$ the charge of the electron. Here
$f(E_{\zeta })=f(E_{\zeta ^{\prime }})$ for elastic scattering and
$X_{\zeta }=\left\langle \zeta \right\vert x\left\vert \zeta \right\rangle$
with $x$ being the position operator.

The scattering rate is given by Fermi's golden rule
\begin{equation}
\hspace*{-0.25cm}W_{\zeta \zeta ^{\prime }}=F 
\sum_{\zeta \neq
\zeta ^{\prime }}\left\vert U(\mathbf{q})\right\vert ^{2}\left\vert J_{\zeta
\zeta ^{\prime }}(u)\right\vert ^{2}\delta (E_{\zeta }-E_{\zeta ^{\prime
}})\delta _{k_{y},k_{y}^{\prime }+q_{y}}, \label{5}
\end{equation}
with $F=2\pi N_i/S_{0}\hslash,\,q^2=q_{x}^{2}+q_{y}^{2}$,
$u=l_{B}^{2}q^2/2$, and $N_i$ the impurity density.
$J_{\zeta \zeta ^{\prime }}(u)=\left\langle \zeta \right\vert \exp
(i{\bf q\cdot r})\left\vert \zeta ^{\prime }\right\rangle $ are the form factors and
$\left\vert \zeta \right\rangle \equiv \left\vert n,s,l,k_{y}\right\rangle $.
$U(\mathbf{q})=U_{0}/(q^{2}+k_{s}^{2})^{1/2}$ with $U_{0}=e^{2}/(2\varepsilon
_{r}\varepsilon _{0})$. Further, $k_{s}$ is the screening wave vector, $\varepsilon
_{r}$ the relative permittivity, and $\varepsilon _{0}$ the permittivity of
the vacuum. Furthermore, if the impurity potential is
short-ranged (of the Dirac $\delta $-function type), one may use the
approximation $k_{s}\gg q$ and obtain $U(\mathbf{q})\approx U_{0}/k_{s}$.
Since the scattering is elastic and the eigenfunctions are degenerate in the quantum
number $k_{x}$, cf.\ Eq.\ (3), only the $n\rightarrow n$ transitions are
allowed. Further, we have $(X_{\zeta}-X_{\zeta ^{\prime }})^{2}=l_{B}^{4}q_{y}^{2}$, 
transform the sums over $k_y$ and $q$ into integrals, and evaluate them using 
cylindrical coordinates. The form factor $\left\vert J_{\zeta \zeta ^{\prime
}}(u)\right\vert ^{2}$ can be evaluated from the matrix element 
$\left\langle \zeta \right\vert \exp
(i{\bf q\cdot r})\left\vert \zeta ^{\prime }\right\rangle$. The result is 
$\left\vert J_{nn}(u)\right\vert ^{2}=\exp (-u)\big( \left\vert C_{n,s}^{\lambda ,l}
\right\vert ^{2}L_{n}(u)+\left\vert D_{n,s}^{\lambda ,l}
\right\vert ^{2}L_{n-1}(u)\big) ^{2}$ for $n=n^{\prime }$. With these details Eq.\ (4) takes the form 
\begin{equation}
\sigma _{xx}=\frac{e^{2}}{h}\frac{N_i\beta U_{0}^{2}}{4u_{sc}\hslash \omega _{c}}
\sum_{s,n }I_{n,s}^{\lambda ,l}\,f(E_{n,s}^{\lambda
,l})\,[1-f(E_{n,s}^{\lambda ,l})], \label{6}
\end{equation}
where $f(E_{n,s}^{\lambda ,l})=(\exp [\beta (\lambda [\hslash
^{2}\omega _{c}^{2}n+(\Delta _{s,l})^{2}]^{1/2}-E_{F})]+1)^{-1}$ and $u_{sc}=l_{B}^{2}k_{s}^2/2$. 
The sum over $s$ is trivial since the two surfaces can be treated 
independently due to the different gaps. The factor
$I_{n,s}^{\lambda ,l}$ in Eq.\ (6) is the integral $\int_{0}^{\infty }u\left\vert J_{nn}(u)\right\vert
^{2}du$ that can be evaluated analytically using the properties of the orthogonal 
polynomials $L_{n}(u)$. The result is
\begin{equation} 
I_{n,s}^{\lambda ,l}=(2n+1)\big\vert C_{n,s}^{\lambda ,l}\big\vert
^{4}-2n\big\vert C_{n,s}^{\lambda ,l}\big\vert ^{2}\big\vert
D_{n,s}^{\lambda ,l}\big\vert ^{2} +(2n-1)\big\vert D_{n,s}^{\lambda ,l}
\big\vert ^{4}. \label{8}
\end{equation} 
For $\Delta _{s,l}=0$, Eq.\ (7) reduces to $2n/4$, which means that the minima of 
$\sigma _{xx}$ occur at the odd factors $\nu =2n+1$ in accord with Ref. \cite{PV}. 

Since the band gap $l \Delta _{\Omega }+s \Delta _{h}$ becomes surface dependent,
see Fig.\ 1, the longitudinal conductivity is dominated by one surface only, that of the
symmetric or antisymmetric surface states.
As usual, this conductivity, given by Eq.\ (6), exhibits Shubnikov-de Haas oscillations. For 
$\Delta _{h}=\Delta _{\Omega }=0$ 
we must consider both surfaces. The electron-hole spectrum is symmetric
 with a single peak (solid curve) at the Dirac point, as shown in Fig.\ 2(a),
using the parameters \cite{DY,HL,CC}: $N_i=1\times 10^{13}$
m$^{-2}$, $\mu _{B}=5.788\times 10^{-5}$ eV/T, $T$ = 2 K, $B=$ 1 T,
 $k_{s}=10^{-7}$ m$^{-1}$, $v_F=5\times 10^{5}$
m/s, and $\epsilon _{r}=4$. We find a gap $\Delta _{h}$ at the Dirac point for
$\Delta _{h}\neq 0$ and $\Delta _{\Omega}=0$ with symmetric electron-hole behaviour (dashed curve). 
This gives $\sigma _{xx}=0$ at the Dirac point
and the peak at $E_F=0$ (solid curve) splits into two peaks, one in the electron
($s=-1$) and one in the hole band ($s=1$) in accord with Eq.\ (2). 

For $\Delta _{\Omega}>\Delta_{h}>0$ the electron-hole spectrum is asymmetric and we 
consider only one surface 
depending on the light's polarization. We consider only the symmetric surface states ($s=1$, black curves) for
left-handed light $(l=-1)$ or the antisymmetric surface states ($s=-1$, red curves) for
right-handed light $(l=1)$ and show $\sigma _{xx}$ in Fig.\ 2(b). 
As seen, the $n=0$ LL shifts into the hole or electron band.
The shift can be understood with the help of the eigenvalues
shown in Fig.\ 1: for right-(left-)handed light the $n=0$ LL moves
into the hole (electron) band. This is a nontrivial state entirely new in FTIs. 
We notice in passing that were we to plot the current polarization 
$P=\big[\sigma_{xx}(l=1)-\sigma_{xx}(l=-1)\big]/\big[\sigma_{xx}(l=1)+\sigma_{xx}(l=-1)\big]$
we would have, on account of Fig. 2(b), only two peaks of height $P=1 (-1)$
centred at $E_F\approx -0.01 (0.01)$ eV.  Also, had we considered the $s=-1$
surface with left-handed light $(l=-1)$ or the $s=1$ surface with right-handed light $(l=1)$, $\sigma_{xx}$
would be zero in the entire range of Fig.\ 2 since the corresponding 
surface states start at $\pm 0.05$ eV, cf.\ Fig.\ 1 for $B=1$ T. This is also
corroborated by the fact that at very low temperatures the factor $\beta f(...)[1-f(...)]$ in Eq.\ (6) behaves as the function
$\delta(E_{n,s}^{\lambda,l} -E_F)$.
\begin{figure}[t]
\includegraphics[width=0.81\columnwidth,height=0.4\columnwidth]{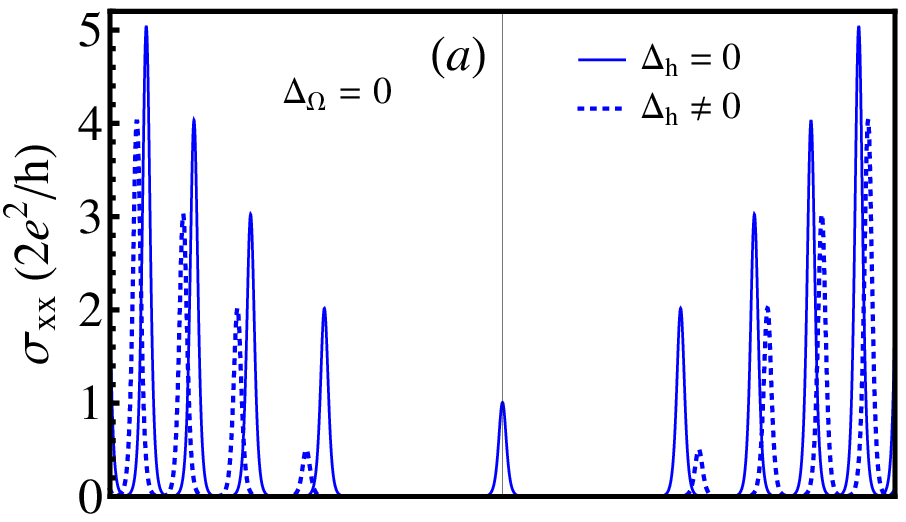}\vspace{-0.4 cm}
\hspace*{0.45 cm}\includegraphics[width=0.85\columnwidth,height=0.5\columnwidth]{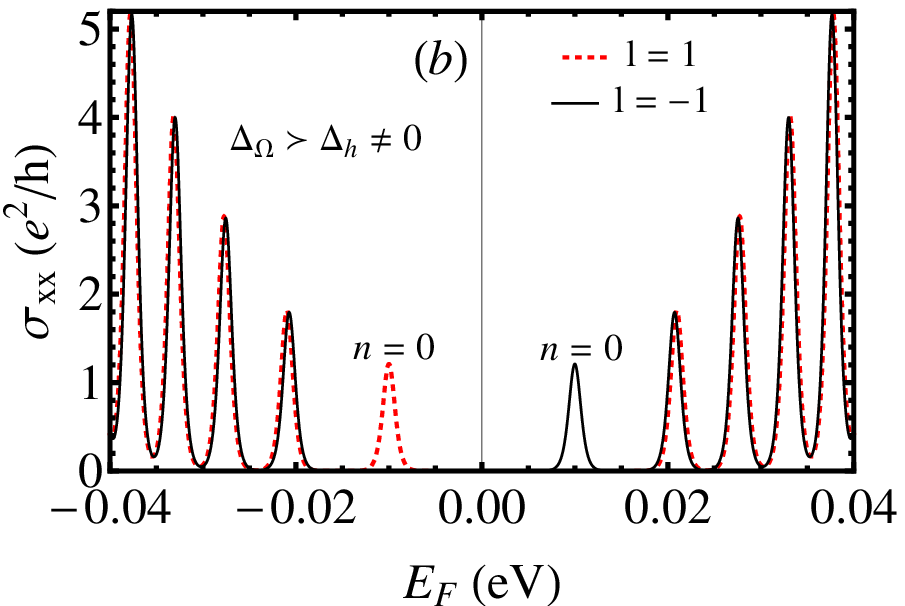}
\vspace{-.25cm}
\caption{Longitudinal conductivity as a function of the Fermi energy $E_F$ for $T=2$ K 
and $B=1$ T. (a) The solid curve is for $\Delta _{\Omega }=\Delta _{h}=0$ meV 
and the dashed one for $\Delta _{h}=20$ meV and $\Delta _{\Omega }=0$ meV.
(b) The red dashed curve is for antisymmetric ($s=-1$) surface states with right-handed
light and the black solid one for symmetric ($s=1$) surface states with left-handed light 
with $\Delta _{h}=20$ meV and $\Delta _{\Omega }=30$ meV.}
\end{figure}

\textit{Hall conductivity}.
For linear responses to a weak source-to-drain
electric field, the Hall conductivity 
is given by the Kubo-Greenwood formula \cite{MK,PV}
\begin{equation}
\sigma _{\mu \nu }
=\frac{i\hslash e^{2}}{S_{0}}\sum_{\zeta \neq \zeta
^{\prime }}\frac{(f_{\zeta }-f_{\zeta ^{\prime }})v_{\nu \zeta \zeta
^{\prime }}v_{\mu \zeta ^{\prime }\zeta }}{(E_{\zeta }-E_{\zeta ^{\prime
}})(E_{\zeta }-E_{\zeta ^{\prime }}+i\Gamma _{\zeta })}, \label{9}
\end{equation}
where $v_{\nu \zeta \zeta ^{\prime }}$ 
and $v_{\mu \zeta ^{\prime }\zeta }$ are the nondiagonal matrix elements of 
the velocity operator with $\mu=x,y, \nu=x,y$. The sum runs over all quantum
numbers of the states $\left\vert \zeta \right\rangle \equiv \left\vert
n,s,l,k_{y}\right\rangle $ and $\left\vert \zeta ^{\prime }\right\rangle
\equiv \left\vert n^{\prime },s^{\prime },l ^{\prime },k_{y}^{\prime
}\right\rangle $ provided $\zeta \neq \zeta ^{\prime }$. Assuming that
the level broadening is approximately the same for all LLs, 
$\Gamma_{\zeta}=\Gamma $, one can show that the imaginary part of Eq.\ (8) vanishes.
To obtain the most transparent results for the Hall conductivity $\sigma _{yx}$,
we take $\Gamma=0$.
The relevant velocity matrix elements are obtained from Eq.\ (1), for $\nu =x$ and $\mu =y$, 
and the evaluation follows  the procedure detailed in Ref.\ \cite{PV}. The result for 
$\sigma _{yx}$ can be expressed as a
sum of two terms, one (I) for $n\geq 1$ and the other (II) for $n=0$, i.e., $\sigma
_{yx}=\sigma _{yx}^{I}+\sigma _{yx}^{II}$, with
\begin{eqnarray}
\hspace*{-0.1cm}\sigma _{yx}^{I} &=&\frac{e^2 }{h}\sum_{s,n=1}^{\infty
}\Big\{(n+1/2)[f_{n,s}^{+,l}-f_{n+1,s}^{+,l}+f_{n,s}^{-,l}-f_{n+1,s}^{-,l}]
\notag \\
&&-{\Delta
_{s,l}\over 2} \,[{f_{n,s}^{+,l}-f_{n,s}^{-,l}\over E_{n,s}^{+,l}} 
-{f_{n+1,s}^{+,l}-f_{n+1,s}^{-,l}\over E_{n+1,s}^{+,l}}]\Big\}. 
\label{13}
\end{eqnarray}
The sum over $n$ starts at $n=1$, because the $n=0$ LL 
is treated separately. That over $s$ is trivial, as 
we consider only one surface at a time. For $n=0$ Eq.\ (8) gives 
\begin{eqnarray}
\sigma _{yx}^{II} &=&\frac{e^{2}}{h}\sum_{s}
\Big\{f_{0,s}^{+,l}+f_{0,s}^{-,l}-(f_{1,s}^{+,l}+f_{1,s}^{-,l})/2 \label{14} \\
&&+\Delta _{s,l}(f_{1,s}^{+,l}-f_{1,s}^{-,l})/2E_{1,s}^{+,l}\Big\}. \notag
\end{eqnarray}
\begin{figure}[t]
\includegraphics[width=0.8\columnwidth,height=0.4\columnwidth]{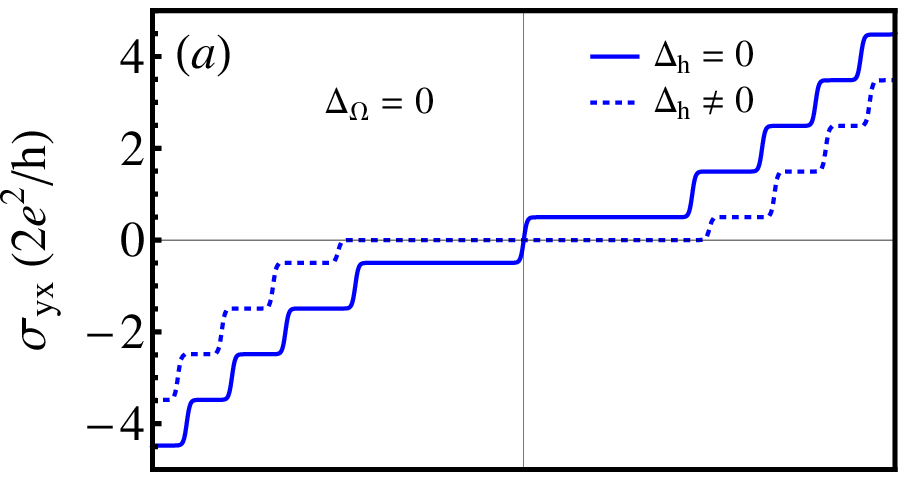}\vspace{-0.4 cm}
\hspace*{0.4 cm}\includegraphics[width=0.85\columnwidth,height=0.5\columnwidth]{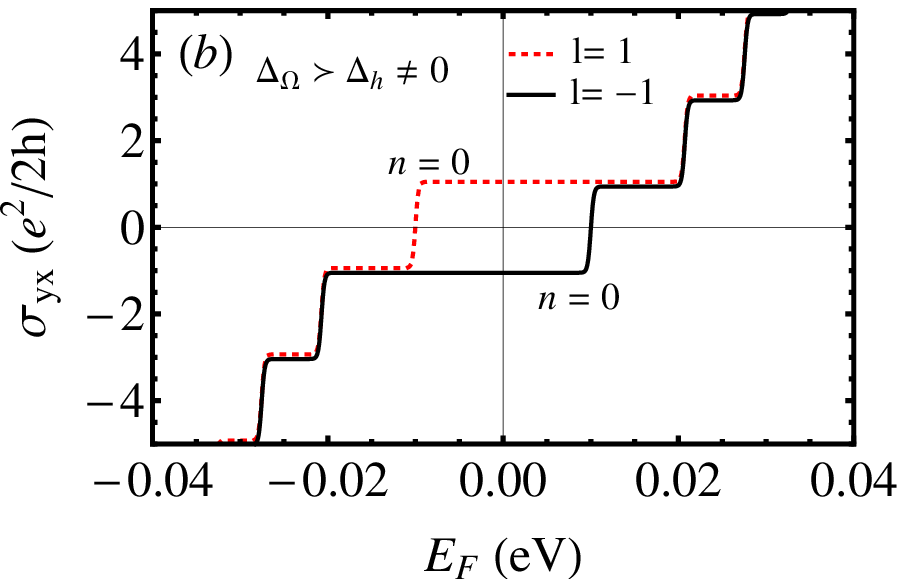}
\vspace{-.25cm}
\caption{ Hall conductivity versus Fermi energy $E_F$. All curves, marked as in Fig.\ 2, are obtained with the same parameters.} 
\end{figure}
At zero or very low temperature the sum over $s$ in Eqs.\ (9) and (10) has
the value $1$ for $n=n_{F}$, since for a single surface the number of filled
states is $1$. 
Here $n_F$ is the LL index at the Fermi energy.
Notice also that for $\Delta _{s,l}=0$ Eqs.\ (9) and (10) 
take the form $\sigma _{yx}=2(e^{2}/h)(n+1/2)$
or in terms of the filling factor $\nu =2(n+1/2)$ as $\sigma _{yx}=\nu e^{2}/h$ for TIs \cite{DY,HL,CC,JG}. 
An important aspect in the QHE in gapless graphene, TIs, silicene, MoS$_2$, etc. 
is the LL at zero energy  or its modification in gapped systems.

Similar to the band structure and longitudinal conductivity, the unconventional QHE is
dominated by one surface. That is, only the symmetric or antisymmetric surface 
contributes to it for left- or right-handed light, respectively.
We show the Hall conductivity $\sigma _{yx}$ in Fig. 3, for $s=1$, as a function of $E_F$.
In the limit $\Delta _{s,l}\to 0$ ((a), solid curve) these results reduce to
an odd-integer QHE in TIs by multiplying with a factor 2 for surface degeneracy \cite{DY,JG}
(limit of zero strain in Ref.\ \cite{CC}), where the plateaus appear
at ($\pm 1,\pm 3,\pm 5,...)e^2/h$, and both surfaces must be considered as in the case of 
Fig.\ 2(a). The results of Fig.\ 3(a), dashed curve, can be reduced to those of single-valley gapped 
graphene for $\Delta _{h}\neq 0$ \cite{PV}, irrespective 
of a factor of 2 due to valley degeneracy, and to those of gapped TIs \cite{CJPC}. 
However, for $\Delta _{\Omega }>\Delta _{h}\neq 0$
only one surface must be considered, see
Fig.\ 2(b) for $\sigma _{xx}$. The Hall plateaus occur at half-integer values
in contrast to previous results for graphene and TIs, as shown in Fig.\ 3 (b). 
At the Dirac point we have $\sigma_{yx}=e^2/2h$ for right-handed light (dashed curve) 
and $\sigma_{yx}=-e^2/2h$ for left-handed light (solid curve), due to the occurrence
of the $n=0$ LL in the hole and electron band, respectively.  Again,
had we considered the $s=-1$ surface with 
left-handed light or the $s=1$ surface with right-handed light, $\sigma_{yx}$ would vanish 
in the $E_F$ range of Fig.\ 3, since the corresponding states would start at
$\pm 0.05$ eV and the occupation factors $f(...)$ would be zero.

The electron-hole symmetry is broken and the plateaus appear at $(\pm 1/2,\pm 3/2,\pm 5/2,...)e^2/h$.
This shows a nontrivial transition at the Dirac point which could be experimentally
tested. It occurs because the energy term $\Delta _{\Omega }$ due to the
\textit{off-resonant} light at the Dirac point can be externally tuned to higher
values \cite{YH,HJJ}. 
As for the influence of level broadening, i.e., finite $\Gamma$, on the results, on
the basis of Ref. \cite{PV} we strongly expect they will not be altered qualitatively.
After all, the QHE has already been realized on TIs \cite{CC}. These signatures of novel quantum phase transitions in FTIs 
are distinct from those in graphene or in TIs without light and relate to different values of the Hall conductivity.
They could be tested in experiments similar to those
performed on semiconducting silicon \cite{DD,JK}. Moreover,  radiation
effects and light-dependent magnetotransport have been
observed for Dirac fermions in the presence of {\it on-resonant} light \cite{PO,JK2}.
Accordingly, we believe our results can be
tested in similar experiments using \textit{off-resonant} light 
\cite{YH,HJJ}, a regime that  is different from that of the
optical absorption spectra \cite{OE,WA}.

\textit{Summary}. We have identified an unconventional QHE in FTIs, in the presence of a 
perpendicular magnetic field, by evaluating
their band structure and the Hall and longitudinal conductivities.
The low-energy dynamics can be governed by a single-surface in a wide range of Fermi energies.
This results in a nontrivial phase transition and unusual Hall plateaus
at {\it half-integer} multiples of $e^{2}/h$ ($\pm 1/2,\pm 3/2,\pm 5/2, ...)$.
In addition, reversing the light polarization
leads to an exchange of surface states, in both the valence and conduction bands, 
and to a shift of the $n=0$ LL into the hole and electron bands, respectively.
These findings suggest new directions in experimental
research and device applications based on FTIs.\\
 
{\bf Acknowledgments}:
This work was supported by the Canadian NSERC Grant No.\ OGP0121756 (MT, PV) 
and by funding from King Abdullah
University of Science and Technology (KAUST) (US).\\[0.2cm]
Electronic addresses: $^{\ddagger}$m.tahir06@alumni.imperial.ac.uk, \\ $^{\dag}$p.vasilopoulos@concordia.ca\\

\end{document}